\documentclass[preprint,showpacs,floats,nofootinbib,superscriptaddress,
  showkeys]{revtex4}

\usepackage{bm}
\usepackage{amsmath}
\usepackage{graphicx}

\begin{document}

\title{The $2p-2h$ electromagnetic response in the quasielastic peak and
  beyond}

\author{A. De Pace}
\author{M. Nardi}
\author{W. M. Alberico}
\affiliation{Istituto Nazionale di Fisica Nucleare, Sezione di Torino and
  Dipartimento di Fisica Teorica, via Giuria 1, I-10125 Torino, Italy}
\author{T. W. Donnelly}
\affiliation{Center for Theoretical Physics, Laboratory for Nuclear Science
  and Department of Physics, Massachusetts Institute of Technology, Cambridge,
  MA 02139, USA}
\author{A. Molinari}
\affiliation{Istituto Nazionale di Fisica Nucleare, Sezione di Torino and
  Dipartimento di Fisica Teorica, via Giuria 1, I-10125 Torino, Italy}

\begin{abstract}
The contribution to the nuclear transverse response function $R_T$
arising from two particle-two hole (2p-2h) states excited through
the action of electromagnetic meson exchange currents (MEC) is
computed in a fully relativistic framework. The MEC considered are
those carried by the pion and by $\Delta$ degrees of freedom, the
latter being viewed as a virtual nucleonic resonance. The
calculation is performed in the relativistic Fermi gas model in
which Lorentz covariance can be maintained. All 2p-2h many-body
diagrams containing two pionic lines that contribute to $R_T$ are
taken into account and the relative impact of the various
components of the MEC on $R_T$ is addressed. The non-relativistic
limit of the MEC contributions is also discussed and compared with
the relativistic results to explore the role played by relativity
in obtaining the 2p-2h nuclear response.
\end{abstract}
\maketitle

\section{Introduction}

Two particle--two hole states in nuclei can be excited through the
action of two-body electromagnetic (EM) currents that enter via
Noether's theorem and the requirement of minimal coupling once an
effective Lagrangian (embodying baryonic and mesonic degrees of
freedom) is invoked. The currents reflect the dual role played by
mesons in nuclear structure studies: mesons carry both the force
which accounts for the binding of nuclei and the currents (in
particular the EM current) which respond to external fields
impinging on the nucleus. Accordingly, the natural classification
into MEC and correlation currents follows. In particular, both
substantially contribute to the inclusive, inelastic scattering of
electrons from nuclei in the excitation energy domain extending
through the quasielastic peak (QEP) to high inelasticity, the
region we shall explore in the present paper.

A long-standing issue of hadronic modeling is the difficulty of
treating currents and interactions consistently, that is, of
respecting gauge invariance by fulfilling the continuity equation.
The model employed in the present study, the relativistic Fermi
gas (RFG) model, has special advantages over most others: gauge
invariance can be respected and fully relativistic modeling, while
not easy, can be undertaken. Given the focus of modern electron
scattering experiments on kinematics where the energies and
momenta transferred to the nucleus are large, and thus where
relativity is expected to be important, one has strong motivation
to explore such models in which fundamental symmetries can be
maintained, even at the expense of accepting their obvious
dynamical limitations. Given the necessity to break the problem
down into tractable pieces, in the present work we forgo the issue
of gauge invariance and focus on a fully relativistic description
of the MEC contribution to the EM excitation of the 2p-2h states
of the RFG. This is already a non-trivial computational problem
and thus the issue of how gauge invariance can be maintained
(which is presently also being explored in other work) is left for
future presentation.

Furthermore, we shall limit our attention in this work to the
roles played by the pion and $\Delta$ in obtaining the MEC
contributions specifically to the transverse response $R_T$. The
relevance of the $\pi$ and $\Delta$ in $R_T$ for the physics of
the quasielastic regime is well-established and, moreover, we wish
to compare our results with the only existing fully relativistic
computation of the 2p-2h MEC contribution to $R_T$, namely that of
\cite{Dek94} (quoted as DBT in the following). That this is to our
knowledge the only previous study of this type is not surprising.
Indeed, to carry out our project it has been necessary to compute
a very large number of terms. Specifically, to get the direct
contribution (see below) we have had to compute about 3000 terms,
whereas to get the exchange contribution (which, as we shall see,
comes into play only via the $\Delta$-isobar) it has been
necessary to compute more than 100,000  terms. The analytic
manipulation of the traces of the relativistic currents has been
done using the algebraic computer program FORM \cite{Ver00}, and
the numerical computation of the individual relativistic
contributions requires up to seven-dimensional Monte Carlo
integrations.

To make direct comparison between our results and
those of DBT we have used exactly the same
form factors and $\Delta$-width of the latter, although
more modern versions of these quantities can straightforwardly be incorporated
--- in future work we will update our predictions by doing so.
In order to appreciate the importance of Lorentz covariance, we
have calculated $R_T$ for the RFG not only fully relativistically, but also
in the non-relativistic limit. For the latter a few calculations are available,
specifically those carried out in \cite{Dek94,Van80,Alb84,Gil97}.
As we shall see, in the current work some important differences between the DBT
results and our results in the high energy domain are found, notably in the
non-relativistic approximations obtained in the two studies.

\section{Two-Body Meson-Exchange Currents in Free Space}

\begin{figure}
\includegraphics[clip,height=8cm]{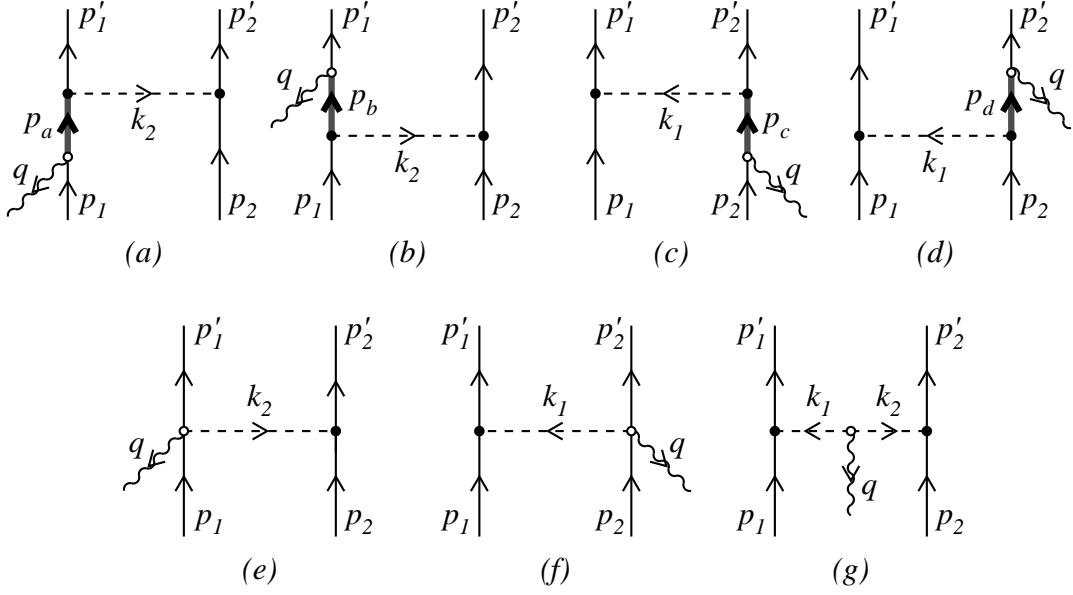}
\caption{\label{fig:currents}The two-body meson exchange currents in
  free-space. The thick lines in the diagrams {\em (a)} to {\em (d)}
represent the $\Delta$ propagation.}
\end{figure}

In Fig.~\ref{fig:currents} we display the free-space two-body
isovector MEC entering in our calculation. Using the labelling in
the figure and defining the four-momenta $k_1=p_1'-p_1$ and
$k_2=p_2'-p_2$ (the four-momentum carried by the virtual photon is
then $q=-k_1-k_2$) their relativistic expression is\footnote{In
this section the current ``operators'' should actually be regarded
as the spin-isospin operator parts of the corresponding matrix
elements between the two-particle states specified in the diagrams
of Fig.~\protect\ref{fig:currents}.}
\begin{equation}
  \label{eq:Jmupi}
  \bm{J}^\mu_f(k_1,k_2) = -i \frac{1}{V^2}
    \frac{f_{\pi NN}^2 f_{\gamma\pi\pi}}{\mu_\pi^2}
    (\bm{\tau}^{(1)}\times\bm{\tau}^{(2)})_3
    \Pi(k_1)_{(1)} \Pi(k_2)_{(2)}
    (k_2-k_1)^\mu
\end{equation}
for the pion-in-flight current (diagram {\em (g)}) and
\begin{equation}
  \label{eq:Jmus}
  \bm{J}^\mu_s(k_1,k_2) = - i \frac{1}{V^2}
    \frac{f_{\pi NN}f_{\gamma\pi NN}}{\mu_\pi^2}
    (\bm{\tau}^{(1)}\times\bm{\tau}^{(2)})_3
    \left[\Pi(k_2)_{(2)}(\gamma^\mu\gamma^5)_{(1)} -
    \Pi(k_1)_{(1)}(\gamma^\mu\gamma^5)_{(2)}\right]
\end{equation}
for the seagull (or contact) current (diagrams {\em (e)} and {\em (f)}).
In the above
\begin{equation}
  \Pi(k)_{(i)} = \frac{\bigl(\rlap/k\gamma^5\bigr)_{(i)}}{k^2-\mu_\pi^2},
\end{equation}
$\mu_\pi$ and $M$ are the pion and nucleon masses, respectively, $V$ is the
(large) volume enclosing the Fermi gas and $f_{\pi NN}$
($f_{\pi NN}^2/4\pi=0.08$) the pseudo-vector pion-nucleon coupling
constant.
The other coupling constants are $f_{\gamma\pi\pi}=1$ and
$f_{\gamma\pi NN}=f_{\pi NN}$; for simplicity, here we omit the form factors
(but they are included in the calculations of the response).
The index $(i)$ attached to the vertex operators distinguishes between
the two interacting nucleons.

The $\Delta$ current is derived from the Peccei pseudo-vector
Lagrangian and reads
\begin{eqnarray}
  \label{eq:JmuD}
  \bm{J}^\mu_\Delta(k_1,k_2) &=& -\frac{1}{V^2}
    \frac{f_{\pi NN}f_{\pi N\Delta}f_{\gamma N\Delta}}{2M\mu_\pi^2}
    \left\{\left[\left(\frac{2}{3}\tau_3^{(2)} -
    \frac{i}{3}(\bm{\tau}^{(1)}\times\bm{\tau}^{(2)})_3\right)
    \left(j^\mu_{(a)}(p_a,k_2,q)\gamma_5\right)_{(1)}\right.\right.\nonumber\\
  && + \left.\left.\left(\frac{2}{3}\tau_3^{(2)} +
    \frac{i}{3}(\bm{\tau}^{(1)}\times\bm{\tau}^{(2)})_3\right)
    \left(\gamma_5 j^\mu_{(b)}(p_b,k_2,q)\right)_{(1)}\right]
    \Pi(k_2)_{(2)} + (1\leftrightarrow2)\right\}.
\end{eqnarray}
It corresponds to diagrams {\em (a)-(d)} of Fig.~\ref{fig:currents}. In the
above equations 
$M_\Delta$ denotes the isobar mass, $p_a\equiv p_1-q$, $p_b\equiv p_1'+q$ and
$f_{\pi N\Delta}=0.54$, $f_{\gamma N\Delta}=5$ yield the strength of the
coupling of the $\Delta$ to the EM and pionic fields,
respectively.

In Eq.~(\ref{eq:JmuD}) the following definitions have been introduced:
\begin{subequations}
\label{eq:jcurr}
\begin{equation}
  j_{(a)\mu}(p,k,q) = (4 k_{\beta} - \rlap/k\gamma_\beta)
    S^{\beta\gamma}(p,M_\Delta) \frac{1}{2}
    \left(-\gamma_\mu \rlap/q \gamma_\gamma + q_\mu\gamma_\gamma\right)
\end{equation}
and
\begin{equation}
  j_{(b)\mu} (p,k,q) = \frac{1}{2}\left(-\gamma_\beta \rlap/q \gamma_\mu
    + q_\mu\gamma_\beta\right) S^{\beta\gamma}(p,M_\Delta)
    (4 k_{\gamma} - \gamma_\gamma \rlap/k);
\end{equation}
\end{subequations}
the analogous expressions
associated with the diagrams (c) and (d)
are simply obtained through the interchange ($1\leftrightarrow2$) as indicated
in Eq.~(\ref{eq:JmuD}), which also implies $p_a\leftrightarrow p_c$
and $p_b\leftrightarrow p_d$ ($p_c\equiv p_2-q$, $p_d\equiv p_2'+q$).

Furthermore, the Rarita-Schwinger (RS) $\Delta$ propagator, namely
\begin{equation}
\label{eq:Deltaprop}
  S^{\beta\gamma}(p,M_\Delta) =
    \frac{\gamma\cdot p + M_\Delta}{p^2-{M_\Delta}^2}
    \left( g^{\beta\gamma} - \frac{\gamma^\beta \gamma^\gamma}{3} -
    \frac{2p^\beta p^\gamma}{3{M_\Delta}^2} -
    \frac{\gamma^\beta p^\gamma - \gamma^\gamma p^\beta}{3M_\Delta}\right),
\end{equation}
 is used (we employ here the metric
$a_\mu b^\mu=a^0b^0-\bm{a}\cdot\bm{b}$).

In the Appendix, the explicit expressions of Eqs.~(\ref{eq:jcurr}a-b) are
reported.

\subsection*{The non-relativistic limit for the currents}

The above given currents are the basic ingredients required in modeling $R_T$.
Since one of our goals is to assess the quality of a given non-relativistic
approximation for $R_T$, we also need the corresponding matrix elements of the
currents in that limit. 
Under the assumption that the momenta of the nucleons involved in the
response are small in comparison to their rest mass (which may or may not be
valid) one obtains\footnote{For brevity, the momentum labels implicitly include
the spin ($\sigma$)  and isospin ($\tau$) quantum numbers.}

\begin{eqnarray}
  J_f(\bm{k}_1,\bm{k}_2) &=& -\frac{(2M)^2}{V^2}
    \frac{f_{\pi NN}^2f_{\gamma\pi\pi}}{\mu_\pi^2}
    4\tau_1\delta_{\tau_1,-\tau_2}\delta_{\tau_1',-\tau_1}
    \delta_{\tau_2',-\tau_2} \frac{\chi^\dagger_{\sigma_1'}
    \bm{\sigma}\cdot\bm{k}_1\chi_{\sigma_1}}{\bm{k}_1^2+\mu_\pi^2} \;
    \frac{\chi^\dagger_{\sigma_2'}\bm{\sigma}\cdot\bm{k}_2\chi_{\sigma_2}}
    {\bm{k}_2^2+\mu_\pi^2} \bigl(\bm{k}_2-\bm{k}_1\bigr) , \nonumber\\
\label{eq:piinflight}
\\
  J_s(\bm{k}_1,\bm{k}_2) &=& -\frac{(2M)^2}{V^2}
    \frac{f_{\pi NN}f_{\gamma\pi NN}}{\mu_\pi^2}
    4\tau_1\delta_{\tau_1,-\tau_2}\delta_{\tau_1',-\tau_1}
    \delta_{\tau_2',-\tau_2} \nonumber \\
  && \qquad\times \left[ \chi^\dagger_{\sigma_1'} \bm{\sigma} \chi_{\sigma_1}
    \frac{\chi^\dagger_{\sigma_2'}\bm{\sigma}\cdot\bm{k}_2\chi_{\sigma_2}}
    {\bm{k}_2^2+\mu_\pi^2}
    -\frac{\chi^\dagger_{\sigma_1'}\bm{\sigma}\cdot\bm{k}_1\chi_{\sigma_1}}
    {\bm{k}_1^2+\mu_\pi^2} \chi^\dagger_{\sigma_2'} \bm{\sigma} \chi_{\sigma_2}
    \right]
\end{eqnarray}
and
\begin{eqnarray}
  J_\Delta (\bm{k}_1,\bm{k}_2) &=&
    \frac{8Mf_{\pi NN}f_{\pi N\Delta}f_{\gamma N\Delta}} {3 V^2 \mu_\pi^2}
    \left\{ \vphantom{\frac{\chi^\dagger_\sigma}{\bm{k}_2^2+\mu_\pi^2}}
    \left[ \vphantom{\chi^\dagger_{\sigma_1'}}
    i B \delta_{\tau_1'\tau_1}\delta_{\tau_2'\tau_2}
    \bigl(\bm{q}\times\bm{k}_2\bigr) \delta_{\sigma_1'\sigma1}
    \right. \right. \nonumber \\
  && \left.\left. - A
    \delta_{\tau_1,-\tau_2}\delta_{\tau_1',-\tau_1}\delta_{\tau_2',-\tau_2}
    \chi^\dagger_{\sigma_1'}\bigl[\bm{q}\times(\bm{k}_2\times\bm{\sigma})\bigr]
    \chi_{\sigma_1}\right] \tau_2
    \frac{\chi^\dagger_{\sigma_2'}\bm{\sigma}\cdot\bm{k}_2\chi_{\sigma_2}}
    {\bm{k}_2^2+\mu_\pi^2} + (1\leftrightarrow 2) \right\}, \nonumber \\
\label{eq:NRdcurr}
\end{eqnarray}
where
\begin{subequations}
\begin{eqnarray}
  A = 2\frac{2M_\Delta+M-\frac{\displaystyle 2M^2}{\displaystyle 3M_\Delta}
    +\frac{\displaystyle M^3}{\displaystyle 3M_\Delta^2}}{(M_\Delta^2-M^2)} =
    2\frac{6M_\Delta^2-3MM_\Delta+M^2}{3M_\Delta^2(M_\Delta-M)}
\end{eqnarray}
and
\begin{eqnarray}
  B = 2\frac{2M_\Delta+3M+\frac{\displaystyle 2M^2}{\displaystyle 3M_\Delta}
    -\frac{\displaystyle M^3}{\displaystyle 3M_\Delta^2}}{(M_\Delta^2-M^2)}
    =2\frac{6M_\Delta^2+3MM_\Delta-M^2}{3M_\Delta^2(M_\Delta-M)}.
\end{eqnarray}
\label{eq:coeffAB}
\end{subequations}

It should be noted, however, that this is not the only possible
choice for a non-relativistic $\Delta$ current. For instance the
one employed in \cite{Van80} and \cite{Alb84} indeed has the same
structure as Eq.~(\ref{eq:NRdcurr}), but occurs with coefficients
given by the following expressions
\begin{eqnarray}
  A=2\frac{(2M_\Delta+M)}{M_\Delta^2-M^2}
\quad\quad\quad \text{and}\quad\quad\quad \
  B=2\frac{(2M_\Delta+3M)}{M_\Delta^2-M^2},
\label{AABB}
\end{eqnarray}
which are obtained by disregarding the last two terms in the numerator of
Eq.~(\ref{eq:coeffAB}a-b). Note that these terms have opposite sign and affect
the magnitude of the coefficients $A$ and $B$, in opposite direction,
by less than 10\%\footnote{Obviously, the expressions in Eqs.~(\ref{AABB}) can
be obtained by using the simplified form for the RS $\Delta$ propagator,
$S^{\beta\gamma}(p,M_\Delta)=
(\rlap/p+M_\Delta)/(p^2-M_\Delta^2)g^{\beta\gamma}$,
 the other terms giving contributions which are
either zero or proportional to $1/M_\Delta$ and $1/M_\Delta^2$.
Note that the second term, namely $-\gamma^\beta\gamma^\gamma/3$,
never contributes to EM processes.}. The non-relativistic
$\Delta$-current employed in DBT is again given by
Eq.~(\ref{eq:NRdcurr}), however with coefficients
\begin{eqnarray}
A=\frac{8}{3(M_\Delta-M)} \quad\quad\quad\quad\text{and}
\quad\quad\quad\quad B=2A,
\end{eqnarray}
derived by setting $M=M_\Delta$ in the numerator of the right hand side of
Eqs.~(\ref{eq:coeffAB}a-b).
The 2p-2h response function is stable with respect to these different
definitions of the coefficients $A$ and $B$ (the changes amounting to a few
percent).

\section{The response function}

We turn now to the 2p-2h transverse response
function, which can be obtained through the hadronic tensor. The latter is
defined according to\footnote{In the previous sections, the nucleon
momenta $p_i$ and $p_i'$ were arbitrary. From now on, we will indicate
with $p_i$ and $p_i'$, respectively, momenta below (holes) and above
(particles) the Fermi momentum $k_F$. The pion momenta have, as before,
$k_i=p_i'-p_i$.}
\begin{eqnarray}
  W_{\mu\nu}(q^2) = \frac{(2\pi)^3V}{4} \frac{V^4}{(2\pi)^{12}}
    && \int \frac{d\bm{p}_1'\,d\bm{p}_2'\,d\bm{p}_1\,d\bm{p}_2}
    {16 E_{\bm{p}_1'}E_{\bm{p}_2'}E_{\bm{p}_1}E_{\bm{p}_2}}
    \delta^{(4)}(q+p_1'+p_2'-p_1-p_2) \nonumber\\
  && \qquad \times \langle F|\bm{J}_\mu^\dagger|p_1'p_2'p_1p_2\rangle
    \langle p_1'p_2'p_1p_2|\bm{J}_\nu|F\rangle,
\label{eq:htens}
\end{eqnarray}
where
$|F\rangle$ represents the Fermi sphere, $E_{\bm{p}}=\sqrt{\bm{p}^2+M^2}$,
$\bm{J}\equiv \bm{J}_\pi+\bm{J}_\Delta = \bm{J}_f+\bm{J}_s+\bm{J}_\Delta$
and the factor 1/4 accounts for
the identity of the 2 particles and 2 holes in the final states.

Being concerned with $R_T$ in the present work, actually we only need the
spatial 
components of $W_{\mu\nu}$ (in fact only the $\mu=\nu=1$ and 2 components in
the combination $11+22$, since the unpolarized transverse response is the
focus). 
That is, we have
\begin{eqnarray}
\label{eq:RT}
  R_T(\bm{q},\omega) &=& \frac{(2\pi)^3V}{4} \frac{V^4}{(2\pi)^{12}}
    \int \frac{d\bm{p}_1'\,d\bm{p}_2'\,d\bm{p}_1\,d\bm{p}_2}
    {16 E_{\bm{p}_1'}E_{\bm{p}_2'}E_{\bm{p}_1}E_{\bm{p}_2}}
    \theta(|\bm{p}_1'|-k_F) \theta(|\bm{p}_2'|-k_F) \theta(k_F-|\bm{p}_1|)
    \nonumber \\
  && \times \theta(k_F-|\bm{p}_2|)
    \delta[\omega-(E_{\bm{p}_1'}+E_{\bm{p}_2'}-E_{\bm{p}_1}-E_{\bm{p}_2})]
    \delta^{(3)}(\bm{q}+\bm{p}_1'+\bm{p}_2'-\bm{p}_1-\bm{p}_2)
    \nonumber \\
  && \times 2 \sum_{\sigma \tau} \sum_{i,j=1}^3
    \left(\delta_{ij}-\frac{q_i q_j}{\bm{q}^2}\right)
    \left[J_i^\dagger(\bm{p}_1'-\bm{p}_1,\bm{p}_2'-\bm{p}_2)
    J_j(\bm{p}_1'-\bm{p}_1,\bm{p}_2'-\bm{p}_2) \nonumber \right. \\
  && \qquad\qquad\qquad\qquad\qquad\left.
    - J_i^\dagger(\bm{p}_1'-\bm{p}_1,\bm{p}_2'-\bm{p}_2)
    J_j(\bm{p}_1'-\bm{p}_2,\bm{p}_2'-\bm{p}_1)\right].
\end{eqnarray}
This expression explicitly displays the direct and exchange
contribution stemming from the antisymmetrization of the 2p-2h final states in
Eq.~(\ref{eq:htens}). In Eq.~(\ref{eq:RT}) the $J_i$ are the matrix elements of
the currents in Eqs.~(\ref{eq:Jmupi}), (\ref{eq:Jmus}) and (\ref{eq:JmuD}) in
momentum space. Note that we define the excitation energy as $\omega=-q_0$.

By eliminating one of the integrations via the $\delta^{(3)}$,
a simpler expression for the transverse response emerges:
\begin{eqnarray}
\label{eq:RTDek}
  R_T(\bm{q},\omega) &=&  \frac{V}{2(2\pi)^9}
    \int {d\bm{p}_1'\,d\bm{p}_1\,d\bm{p}_2}
    \theta(|\bm{p}_1'|-k_F) \theta(|\bm{p}_2'|-k_F) \theta(k_F-|\bm{p}_1|)
    \theta(k_F-|\bm{p}_2|) \nonumber \\
  && \times
    \delta[\omega-(E_{\bm{p}_1'}+E_{\bm{p}_2'}-E_{\bm{p}_1}-E_{\bm{p}_2})]
    \sum_{\sigma \tau} \sum_{i,j=1}^3
    \frac{V^4}{16 E_{\bm{p}_1'}E_{\bm{p}_2'}E_{\bm{p}_1}E_{\bm{p}_2}}
    \left(\delta_{ij}-\frac{q_i q_j}{\bm{q}^2}\right) \nonumber \\
  && \times
    \left[J_i^\dagger(\bm{k}_1,\bm{k}_2) J_j(\bm{k}_1,\bm{k}_2)
    -J_i^\dagger(\bm{k}_1,\bm{k}_2) J_j(\bm{k}_1',\bm{k}_2')\right],
\end{eqnarray}
where now $\bm{p}_2'=-\bm{q}+\bm{p}_1+\bm{p}_2-\bm{p}_1'$.
The exchange term in Eq.~(\ref{eq:RTDek}) depends on the pionic
momenta $\bm{k}_i'$ defined as $\bm{k}_1'=\bm{p}_1'-\bm{p}_2$,
$\bm{k}_2'=\bm{p}_2'-\bm{p}_1$.

By exploiting the remaining $\delta$-function and with some
algebra, the integral defining the transverse response can be
reduced to seven dimensions. In previous work where only the
direct term and the non-relativistic limit were considered, it
turned out to be possible to reduce the integration to a
bi-dimensional one; but now for the relativistic expression, this
is no longer possible, even for the direct term.

As anticipated in the Introduction the traces of the relativistic currents
typically generate a huge number of terms, so that it is not practical to
report explicitly the expressions used in the numerical calculations of the
relativistic responses. However, in the Appendix we give the formulae for the
direct pion and pion/$\Delta$ contributions, the only ones having sizes
suitable for publication.

Numerical integrations have been performed using Monte Carlo techniques,
varying the sample size until a standard deviation better than 1\% is
reached. 
The number of configurations required for such an accuracy is $10^6\div10^7$.
A more stringent test of the accuracy in the numerical calculations can be
obtained in the case of the direct non-relativistic contribution, since this 
can be reduced to a two-dimensional integral \cite{Van80} and done through
standard quadrature. Also here the agreement is at the 1\% level. 

\subsection*{The non-relativistic limit}

Before addressing the issue of the numerical evaluation of the
fully relativistic problem, below we report the non-relativistic
limit of the integrands in Eq.~(\ref{eq:RTDek}), to be referred to
as ${\cal R}_T^D(\bm{k}_1,\bm{k}_2;\bm{q},\omega)$ and ${\cal
R}_T^E(\bm{k}_1,\bm{k}_2;\bm{k}'_1,\bm{k}'_2;\bm{q},\omega)$, for
the direct and exchange contributions, respectively. This  will
allow us to connect the various pieces contributing to the
transverse response to specific Feynman diagrams and to ascertain
from whence the major contributions arise. While the relativistic
expressions for the transverse response function are quite
cumbersome, the non-relativistic limit provides a relatively
simple and controllable environment in which to check the
correctness of the calculations. We also remark that our
expression for ${\cal R}_T^D$ will turn out to differ somewhat
from the one derived in DBT (no explicit formula is given there
for the exchange part).

\begin{figure}
\includegraphics[clip,width=16cm]{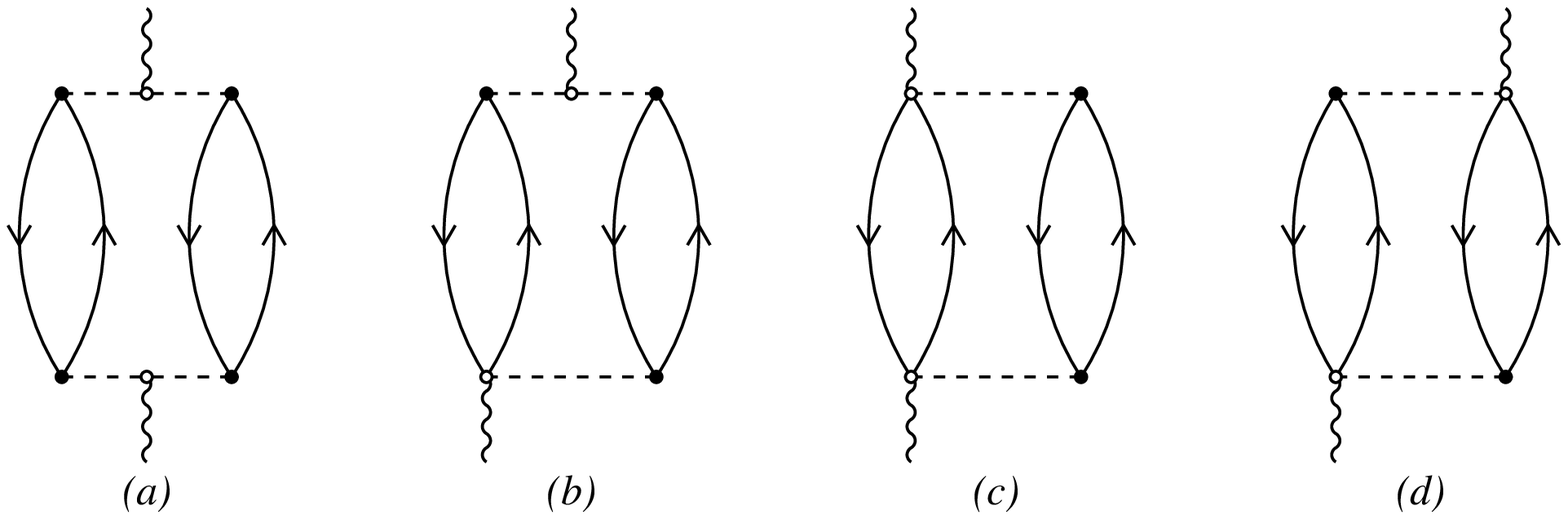}
\caption{\label{fig:MECpion}The direct pionic contributions to the MEC 2p-2h
  response function.}
\includegraphics[clip,width=16cm]{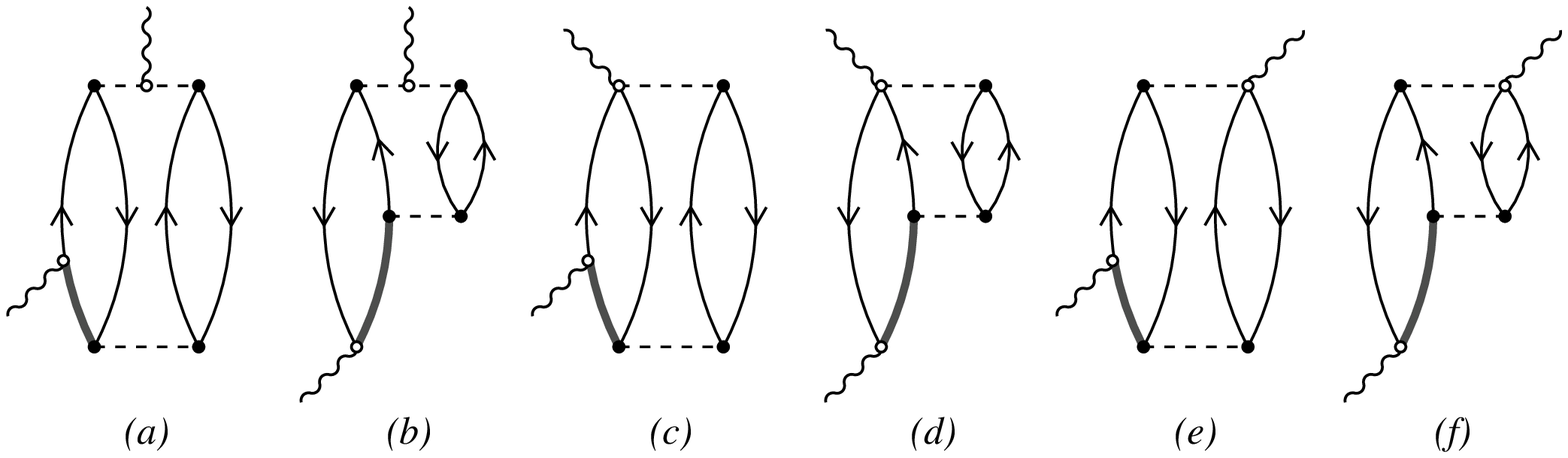}
\caption{\label{fig:MECint}The direct pionic/$\Delta$ interference
  contributions to the MEC 2p-2h response function.}
\includegraphics[clip,width=16cm]{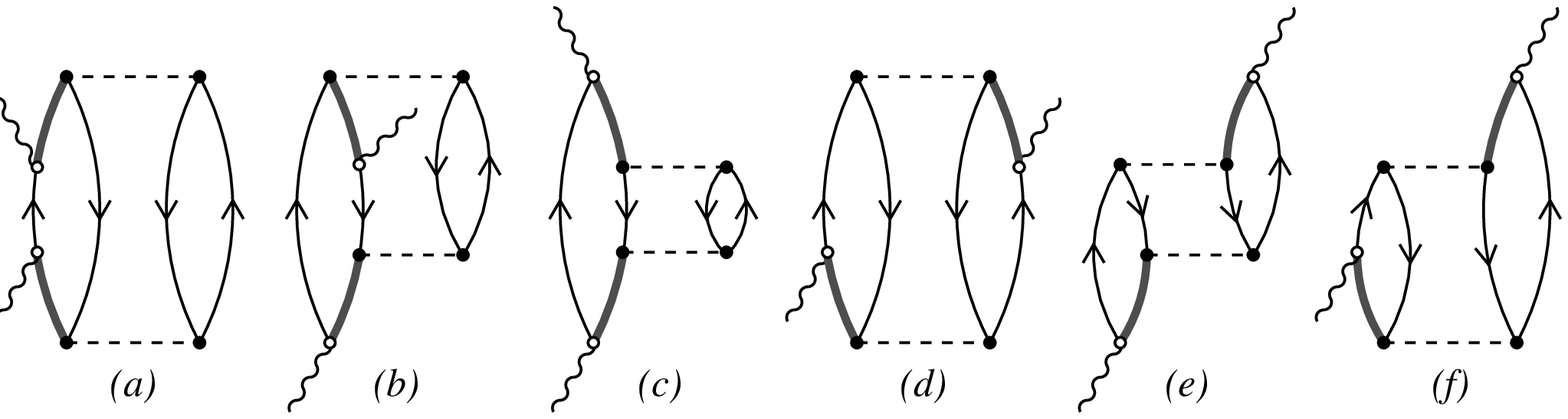}
\caption{\label{fig:MECDelta}The direct $\Delta$ contributions to the MEC 2p-2h
  response function.}
\end{figure}

For the purely pionic contribution and
for the interference between the pionic and $\Delta$ current we get,
respectively, 
\begin{eqnarray}
\label{eq:MECpion}
  {\cal R}_T^{D\pi}(\bm{k}_1,\bm{k}_2;\bm{q},\omega) &=&
    \frac{V^4}{(2M)^4} \sum_{\sigma\tau}
    \sum_{ij} \left(\delta_{ij}-\frac{q_i q_j}{\bm{q}^2}\right)
    J_i^{\pi\dagger}(\bm{p}_1'-\bm{p_1},\bm{p}_2'-\bm{p}_2)
    J_j^{\pi}(\bm{p}_1'-\bm{p_1},\bm{p}_2'-\bm{p}_2) \nonumber \\
  &=& 128 F^2_{\gamma NN}(q^2)
    \left\{ \frac{f_{\pi NN}^4f_{\gamma\pi\pi}^2}{\mu_\pi^4}
    \frac{\bm{k}_{1T}^2 \,\bm{k}_1^2\bm{k}_2^2F_{\pi NN}^2(\bm{k}_1^2) F_{\pi
    NN}^2(\bm{k}_2^2)}{(\bm{k}_1^2+\mu_\pi^2)^2(\bm{k}_2^2+\mu_\pi^2)^2}
    \right. \nonumber \\
  && \left. \hspace*{-15mm}
    + \frac{f_{\pi NN}^2f_{\gamma\pi NN}^2}{2\mu_\pi^4} \left[
    \frac{\bm{k}_1^2F_{\pi NN}^4(\bm{k}_1^2)}{(\bm{k}_1^2+\mu_\pi^2)^2} +
    \frac{\bm{k}_2^2F_{\pi NN}^4(\bm{k}_2^2)}{(\bm{k}_2^2+\mu_\pi^2)^2} +
    \frac{(\bm{k}_{1T}\!\cdot \! \bm{k}_{2T}) \, F_{\pi NN}^2(\bm{k}_1^2)F_{\pi
    NN}^2(\bm{k}_2^2)}{(\bm{k}_1^2+\mu_\pi^2)(\bm{k}_2^2+\mu_\pi^2)}\right]
   \right. \nonumber \\
  && \left. \hspace*{-15mm}
    - \frac{f_{\pi NN}^3f_{\gamma\pi NN}f_{\gamma\pi\pi}}{\mu_\pi^4} \left(
    \frac{\bm{k}_{1T}^2\bm{k}_1^2F_{\pi NN}^3(\bm{k}_1^2) F_{\pi
    NN}(\bm{k}_2^2)}{(\bm{k}_1^2+\mu_\pi^2)^2(\bm{k}_2^2+\mu_\pi^2)}
    +\frac{\bm{k}_{2T}^2\bm{k}_2^2F_{\pi NN}^3(\bm{k}_2^2) F_{\pi
    NN}(\bm{k}_1^2)}{(\bm{k}_1^2+\mu_\pi^2)(\bm{k}_2^2+\mu_\pi^2)^2}
    \right) \right\} \nonumber \\
\end{eqnarray}
and
\begin{eqnarray}
\label{eq:MECint}
  {\cal R}_T^{D \{ \pi\Delta\} }(\bm{k}_1,\bm{k}_2;\bm{q},\omega) &=&
    \frac{V^4}{(2M)^4} \sum_{\sigma\tau}
    \sum_{ij} \left(\delta_{ij}-\frac{q_i q_j}{\bm{q}^2}\right) \nonumber \\
  && \times \Bigl[
    J_i^{\pi\dagger}(\bm{k}_1,\bm{k}_2) J_j^{\Delta}(\bm{k}_1,\bm{k}_2)
    + J_i^{\Delta\dagger}(\bm{k}_1,\bm{k}_2) J_j^{\pi}(\bm{k}_1,\bm{k}_2)
    \Bigr] \nonumber \\
  &=& 32 A \frac{f_{\pi NN}^2f_{\pi N\Delta}f_{\gamma N\Delta}}
    {3\mu_\pi^2 M} F_{\gamma NN}(q^2)F_{\gamma N\Delta}(q^2) \nonumber \\
  && \times \left\{ \frac{f_{\pi NN}f_{\gamma\pi\pi}}{\mu_\pi^2}
    F_{\pi NN}^2(\bm{k}_1^2)F_{\pi NN}(\bm{k}_2^2)
    F_{\pi N\Delta}(\bm{k}_1^2) \frac{2\bm{q}^2\bm{k}_1^2\bm{k}_{1T}^2}
    {(\bm{k}_1^2+\mu_\pi^2)^2(\bm{k}_2^2+\mu_\pi^2)} \right. \nonumber \\
  && + \frac{f_{\gamma\pi NN}}{\mu_\pi^2}\left[
    F_{\pi NN}^3(\bm{k}_1^2)F_{\pi N\Delta}(\bm{k}_1^2)
    \frac{2\bm{k}_1^2 (\bm{k}_{1}\!\cdot\!\bm{q})}
    {(\bm{k}_1^2+\mu_\pi^2)^2}\right. \nonumber \\
  && \left.\left. + F_{\pi NN}^2(\bm{k}_1^2)F_{\pi NN}(\bm{k}_2^2)
    F_{\pi N\Delta}(\bm{k}_2^2)
    \frac{\bm{q}^2\bm{k}_{1T}^2}{(\bm{k}_1^2+\mu_\pi^2)(\bm{k}_2^2+\mu_\pi^2)}
    \right] + (1\leftrightarrow2)\right\}.  \nonumber \\
\end{eqnarray}
From the different coupling constants in Eqs.~(\ref{eq:MECpion}) and
(\ref{eq:MECint}) it is easy to identify the contributions arising
from the pion-in-flight and seagull currents.
The topologically different diagrams are shown in Figs.~\ref{fig:MECpion}
and \ref{fig:MECint}.

Finally, for the contribution to ${\cal R}_T^D$ of the $\Delta$ current alone,
one has
\begin{eqnarray}
\label{eq:MECDelta}
  {\cal R}_T^{D\Delta}(\bm{k}_1,\bm{k}_2;\bm{q},\omega) &=&
    \frac{V^4}{(2M)^4} \sum_{\sigma\tau}
    \sum_{ij} \left(\delta_{ij}-\frac{q_i q_j}{\bm{q}^2}\right) \nonumber \\
  && \qquad \times {J}_i^{\Delta\dagger}(\bm{p}_1'-\bm{p_1},\bm{p}_2'-\bm{p}_2)
    {J}_j^{\Delta}(\bm{p}_1'-\bm{p_1},\bm{p}_2'-\bm{p}_2) \nonumber \\
  &=& \frac{8f_{\pi NN}^2f_{\pi N\Delta}^2f_{\gamma N\Delta}^2}{9M^2\mu_\pi^4}
    F_{\gamma N\Delta}^2(q^2)\bm{q}^2 \nonumber \\
  && \qquad \times \left\{ \left[
    F_{\pi NN}^2(\bm{k}_1^2)F_{\pi N\Delta}^2(\bm{k}_1^2) \left( B^2
    \frac{2 \bm{k}_{1T}^2 \bm{k}_1^2}{(\bm{k}_1^2+\mu_\pi^2)^2} + A^2
    \frac{(\bm{k}_1^2+\bm{k}_{1L}^2)\bm{k}_1^2}{(\bm{k}_1^2+\mu_\pi^2)^2}
    \right) \right.\right. \nonumber \\
  && \left.\left. \qquad + F_{\pi NN}(\bm{k}_1^2)F_{\pi N\Delta}(\bm{k}_1^2)
    F_{\pi NN}(\bm{k}_2^2)F_{\pi N\Delta}(\bm{k}_2^2) A^2
    \frac{\bm{k}_{1T}^2\bm{q}^2}{(\bm{k}_1^2+\mu_\pi^2)(\bm{k}_2^2+\mu_\pi^2)}
    \right] \right. \nonumber \\
  && \left.\phantom{\frac{\bm{k}_{1T}^2\bm{q}^2}{(\bm{k}_1^2+\mu_\pi^2)}}
    + (1\leftrightarrow2) \right\} ,
\end{eqnarray}
where the first two terms on the right-hand side correspond to the
diagrams {\em (a)--(c)} of Fig.~\ref{fig:MECDelta}, and the last
one to the diagrams {\em (d)--(f)}. In this case six distinct
diagrams contribute.

In Eqs.~(\ref{eq:MECpion}), (\ref{eq:MECint}) and (\ref{eq:MECDelta})
$\bm{k}_L$ and $\bm{k}_T$ indicate the longitudinal and transverse
components of the vector $\bm{k}$ with respect to the direction fixed by
$\bm{q}$.
Furthermore, in the appropriate places, the hadronic monopole form factors
\begin{subequations}
\begin{eqnarray}
  F_{\pi NN}(k^2) &=& \frac{\Lambda_\pi^2-\mu_\pi^2}{\Lambda_\pi^2-k^2}, \\
  F_{\pi N\Delta}(k^2) &=& \frac{\Lambda_{\pi N\Delta}^2}{\Lambda_{\pi
    N\Delta}^2-k^2}
\end{eqnarray}
and the EM ones
\begin{eqnarray}
  F_{\gamma NN}(q^2) &=& \frac{1}{(1-q^2/\Lambda_D^2)^2}, \\
  F_{\gamma N\Delta}(q^2) &=& F_{\gamma NN}(q^2)
    \left(1-\frac{q^2}{\Lambda_2^2}\right)^{-\frac{1}{2}}
    \left(1-\frac{q^2}{\Lambda_3^2}\right)^{-\frac{1}{2}}
\end{eqnarray}
\end{subequations}
have been introduced. In the non-relativistic expressions the
hadronic form factors have been taken in the static limit. The
cut-offs have been chosen as in DBT, namely
$\Lambda_\pi=1300$~MeV, $\Lambda_{\pi N\Delta}=1150$~MeV,
$\Lambda_D^2=0.71$~GeV$^2$, $\Lambda_2=M+M_\Delta$ and
$\Lambda_3^2=3.5$~GeV$^2$. This choice clearly makes it possible a
direct comparison between our results for $R_T$ and those of DBT.

\begin{figure}
\includegraphics[clip,width=16cm]{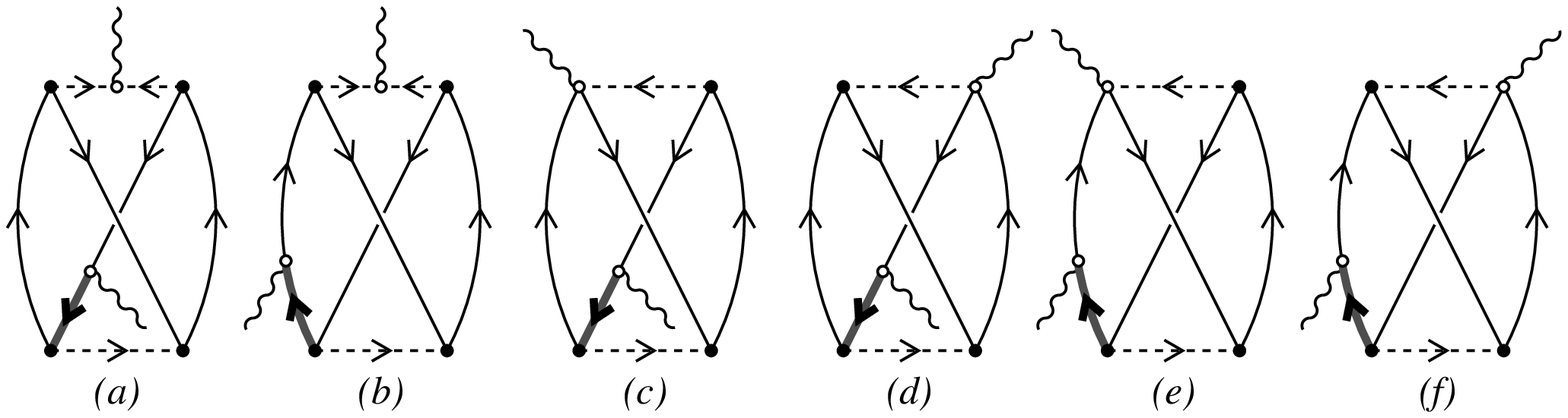}
\caption{\label{fig:DPex}The exchange pionic/$\Delta$ interference
  contributions to the MEC 2p-2h response function.}
\includegraphics[clip,width=16cm]{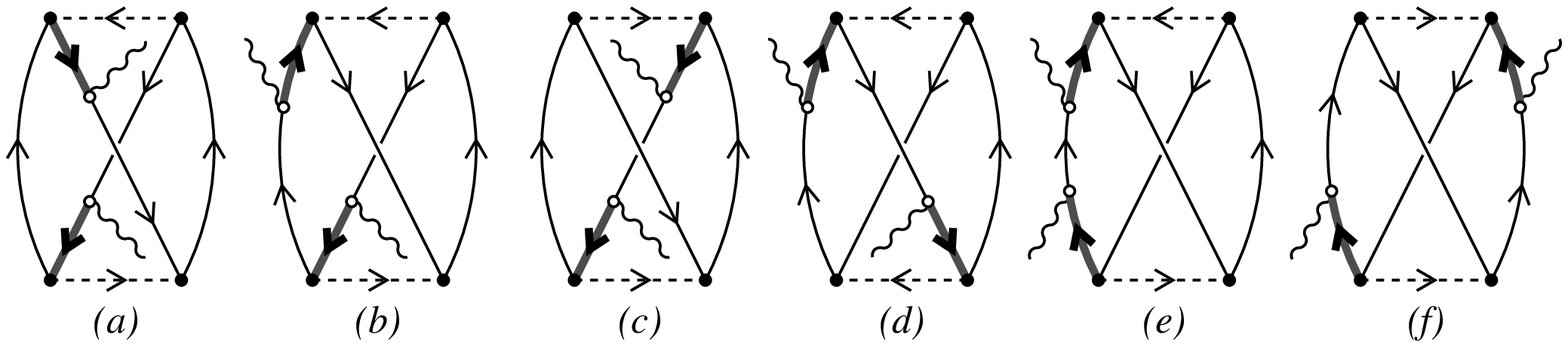}
\caption{\label{fig:DDex}The exchange $\Delta$ contributions to the MEC 2p-2h
  response function.}
\end{figure}

For completeness, we give also the formulae of the (smaller) exchange
contributions to the integrand of Eq.~(\ref{eq:RTDek}),
${\cal R}_T^E(\bm{k}_1,\bm{k}_2;\bm{k}'_1,\bm{k}'_2;\bm{q},\omega)$, in the
non-relativistic limit.
The purely pionic contribution is identically zero, as a consequence of
charge conservation and of the fact that the photon does not
couple to a neutral pion.
For the interference between pion and $\Delta$ (Fig.~\ref{fig:DPex}) we have
\begin{eqnarray}
\label{eq:MECintex}
  && {\cal R}_T^{E \{ \pi\Delta\} }(\bm{k}_1,\bm{k}_2;
    \bm{k}_1',\bm{k}_2';\bm{q},\omega) = \nonumber \\
  && = \frac{V^4}{(2M)^4} \sum_{\sigma\tau}
    \sum_{ij} \left(\delta_{ij}-\frac{q_i q_j}{\bm{q}^2}\right) \Bigl[
    J_i^{\pi\dagger}(\bm{k}_1,\bm{k}_2)J_j^{\Delta}(\bm{k}_1',\bm{k}_2')
    + J_i^{\Delta\dagger}(\bm{k}_1,\bm{k}_2)J_j^{\pi}(\bm{k}_1',\bm{k}_2')
    \Bigr] \nonumber\\
  && = \frac{16 f_{\pi NN}^3f_{\gamma\pi\pi}f_{\gamma N \Delta}
    f_{\pi N\Delta}}{3\mu_\pi^4 M } B \bm{q}^2 \left\{
    \frac{(\bm{k}_2\times \bm{k}_2')_L^2}
    {(\bm{k}_2^2+\mu_\pi^2)(\bm{k}_2'^2+\mu_\pi^2)}
    \left[\frac{1}{\bm{k}_1^2+\mu_\pi^2}+\frac{1}{\bm{k}_1'^2+\mu_\pi^2}\right]
    + (1\leftrightarrow 2) \right\} \nonumber \\
  && + \frac{8 f_{\pi NN}^3f_{\gamma\pi NN}f_{\gamma N \Delta}
    f_{\pi N\Delta}}{3\mu_\pi^4 M } B \left\{
    \frac{(\bm{q}\cdot\bm{k}_{2})\bm{k}_2'^2+(\bm{q}\cdot\bm{k}_{2}')\bm{k}_2^2
    -(\bm{q}\cdot\bm{k}_{2}')(\bm{k}_2\cdot\bm{k}_2')
    -(\bm{q}\cdot\bm{k}_{2})(\bm{k}_2\cdot\bm{k}_2')}
    {(\bm{k}_2^2+\mu_\pi^2)(\bm{k}_2'^2+\mu_\pi^2)} \right. \nonumber\\
  && \qquad \left.
    + \frac{(\bm{q}\!\cdot\!\bm{k}_{1})\bm{k}_2'^2-
    (\bm{q}\cdot\bm{k}_{2}')(\bm{k}_1\cdot\bm{k}_2')}
    {(\bm{k}_1^2+\mu_\pi^2)(\bm{k}_2'^2+\mu_\pi^2)}
    +\frac{(\bm{q}\cdot\bm{k}_{1}')\bm{k}_2^2
    -(\bm{q}\cdot\bm{k}_{2})(\bm{k}_1'\cdot\bm{k}_2)}
    {(\bm{k}_1'^2+\mu_\pi^2)(\bm{k}_2^2+\mu_\pi^2)}
    + (1\leftrightarrow 2)\right\}.
\end{eqnarray}

The contribution of the $\Delta$ alone (Fig.~\ref{fig:DDex}) is instead
\begin{eqnarray}
\label{eq:MECDeltaex}
  && {\cal R}_T^{E\Delta }(\bm{k}_1,\bm{k}_2;
    \bm{k}_1',\bm{k}_2';\bm{q},\omega) = \frac{V^4}{(2M)^4} \sum_{\sigma\tau}
    \sum_{ij} \left(\delta_{ij}-\frac{q_i q_j}{\bm{q}^2}\right)
    J_i^{\pi\dagger}(\bm{k}_1,\bm{k}_2)J_j^{\Delta}(\bm{k}_1',\bm{k}_2')
    \nonumber \\
  && = \frac{4f_{\pi NN}^2 f_{\pi N \Delta}^2 f_{\gamma N\Delta}^2}
    {9M^2\mu_\pi^4} \bm{q}^2 \left\{ B^2 \left[
    \frac{(\bm{k}_1\cdot\bm{k}_1')(\bm{k}_{1T}\cdot\bm{k}_{1T}')}
    {(\bm{k}_1^2+\mu_\pi^2)(\bm{k}_1'^2+\mu_\pi^2)} +
    \frac{(\bm{k}_1\cdot\bm{k}_2')(\bm{k}_{1T}\cdot\bm{k}_{2T}')}
    {(\bm{k}_1^2+\mu_\pi^2)(\bm{k}_2'^2+\mu_\pi^2)}+
    (1\leftrightarrow 2) \right]\right. \nonumber \\
  && \left. \qquad\qquad + AB \left[\frac{2(\bm{k}_1\times\bm{k}_1')_L^2
    -2 k_{1L}k_{1L}'(\bm{k}_1\cdot\bm{k}_1')
    +k_{1L}'^2\bm{k}_1^2+k_{1L}^2\bm{k}_1'^2}
    {(\bm{k}_1^2+\mu_\pi^2)(\bm{k}_1'^2+\mu_\pi^2)}
    \right.\right. \nonumber \\
  && \left. \left. \qquad\qquad \qquad\qquad
    \frac{2(\bm{k}_1\times\bm{k}_2')_L^2
    -2 k_{1L}{k}_{2L}'(\bm{k}_1\cdot\bm{k}_2')
    +k_{2L}'^2\bm{k}_1^2+k_{1L}^2\bm{k}_2'^2}
    {(\bm{k}_1^2+\mu_\pi^2)(\bm{k}_2'^2+\mu_\pi^2)}+
    (1\leftrightarrow 2) \right] \right\}.
\end{eqnarray}

Equations~(\ref{eq:MECpion}), (\ref{eq:MECint}) and
(\ref{eq:MECDelta}) could in principle be compared with Eq.~(5.11)
of DBT; however, the overall normalization of the latter is not
correct, since its dimension is not consistent with its definition
(namely of being the transverse part of the amplitude ${\cal T}$
given in Eq.~(4.8) of DBT); moreover, the relative weights of the
interference and $\Delta$ contributions with respect to the pionic
one differ, in our calculations, by a factor 2 and 4,
respectively, from those of Eq.~(5.11) of DBT.
These factors, however, are not able to explain the marked difference
between our results and those in that paper.
Note that although the authors of DBT
write down exactly the same expressions as we do for the
non-relativistic MEC currents, actually they state that the
non-relativistic procedure to get their Eq.~(5.11) is applied at
the level of the hadronic tensor, that is by reducing the
(cumbersome) exact relativistic response.

\begin{figure}
\includegraphics[clip,height=5cm]{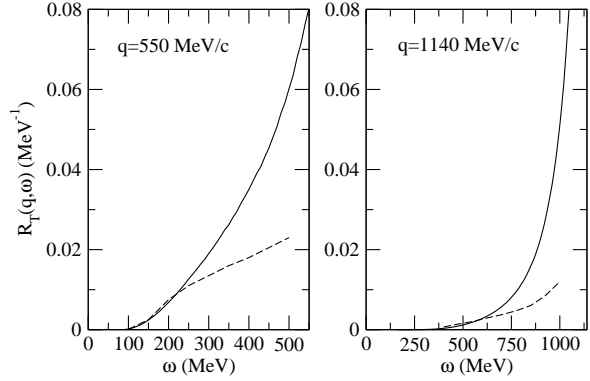}
\caption{\label{fig:RTnr} The non-relativistic transverse response function
  $R_T(q,\omega)$ according to the present calculation (solid) and to the
  one of DBT (dashed) for $q=550$~MeV/c and
  $q=1140$~MeV/c; in both instances $k_F=1.3$~fm$^{-1}$ and
  $\bar{\epsilon}_2=70$~MeV. Only the direct contribution is shown.}
\end{figure}

In Fig.~\ref{fig:RTnr} we now compare our results with those of
DBT, where the non-relativistic $R_T$ (without the exchange
contribution) is shown  for $q=550$~MeV/c (left) and for
$q=1140$~MeV/c (right), with an atomic mass number of 56 and
utilizing a Fermi momentum $k_F=1.3$~fm$^{-1}$. The latter value
is employed for the sake of comparison with DBT, although in fact
it is more appropriate for heavier nuclei. 

It is clearly apparent in the figure that our predictions differ
significantly from those of DBT: while the discrepancy is mild for
moderate values of $\omega$ (roughly, those encompassing the QEP),
it becomes striking at higher energies, namely in the region of
the so-called dip and of the $\Delta$-peak. Here our transverse
response function in the proximity of the lightcone turns out to
be larger by about a factor two at $q=550$~MeV/c and by over a
factor three at $q=1140$~MeV/c.

Note that, in order to conform as closely as possible with the DBT
approach, we have accounted for the initial state binding of the
two holes by phenomenologically inserting a 70~MeV energy shift
$\bar{\epsilon}_2$ in the energy conserving $\delta$-function
appearing in the response function\footnote{This amounts to
replacing $E_{\bm{p}_1}+E_{\bm{p}_2}$ by
$E_{\bm{p}_1}+E_{\bm{p}_2}-\bar{\epsilon}_2$.}. In a
non--relativistic framework this binding energy merely produces a
shift of $R_T$ toward higher $\omega$ values. In the same spirit
and to comply with gauge invariance, we have added (as in DBT) to
the pion-in-flight current in Eq.~(\ref{eq:piinflight}) two terms,
to be viewed as the coupling of the virtual photon to a fictitious
particle of mass equal to the $\pi NN$ cutoff, $\Lambda_{\pi}$, in
spite of the very minor role these terms play in shaping the
transverse response.

Furthermore, it is worth also remarking that our present
calculations agree with the previous findings of  ~\cite{Van80,Alb84}.
On the other hand, the computation of DBT \textit{does not} agree with a
previous one presented by the same authors in ~\cite{Dek92},
notwithstanding that both of them employ the same set of values for the
parameters and the kinematic conditions. Indeed, the former is lower than the
latter by about 25\%.

In the light of these new results, we anticipate contributions
from MEC effects in the region beyond the QEP of a magnitude not
previously foreseen in most previous work. Because the studies of
~\cite{Van80,Alb84} were applied to moderate momenta and to a
restricted range of transferred energies (not exceeding 300~MeV,
in fact) owing to their non-relativistic nature, it is only in the
studies of DBT that suggestions for such large effects can be
found.

\begin{figure}
\includegraphics[clip,height=5cm]{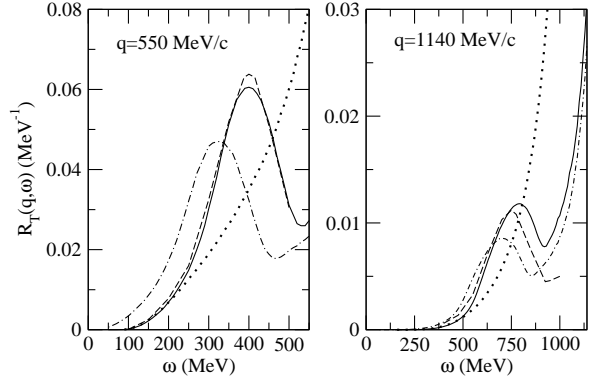}
\caption{\label{fig:RTr} The relativistic transverse response function
  $R_T(q,\omega)$ at $q=550$~MeV/c and $q=1140$~MeV/c calculated with
  $\bar{\epsilon}_2=70$~MeV (solid) and with $\bar{\epsilon}_2=0$
  (dot-dashed). Only the direct contribution is shown.
  The non-relativistic results are also displayed in order to shed light on
  the role of relativity in the response (dotted). For the sake of comparison
  the relativistic results obtained in DBT are displayed (dashed).
  In all instances $k_F=1.3$~fm$^{-1}$.}
\end{figure}

In the next section we shall explore whether or not the fully relativistic
treatment and  the inclusion of the exchange terms modify this finding.

\section{Results in the relativistic regime}

\subsection*{Direct Contributions}

In Fig.~\ref{fig:RTr} we display our relativistic results both
without and with the energy shift $\bar{\epsilon}_2$, for the same
conditions as those of Fig.~\ref{fig:RTnr}. The exchange
contribution is neglected here as well as in all the other figures
of this subsection and considered separately below. In
Fig.~\ref{fig:RTr} we also compare our relativistic predictions
both with those of DBT and with our non-relativistic results.

From the figure we observe the following.
\begin{description}
\item{a)} The fully relativistic calculation differs considerably under some
  circumstances from the fully non-relativistic approximation (i.~e., the solid
  and dotted curves in Fig.~\ref{fig:RTr}, respectively).
  Specifically relativity implies an increase of $R_T$ over the
  non-relativistic 
  results in the region up to the $\Delta$ peak. Specifically, the overall
  effects are small in the domain of the QEP, are modest in the dip region and
  are substantial in the region of the $\Delta$-peak.
  In the region beyond the $\Delta$ peak, relativity instead yields a
  substantial reduction of the response with respect to the non-relativistic
  predictions.
  Of course, a hybrid approach can also be adopted in which spinor matrix
  elements and kinematics are kept non-relativistic, but the dynamic $\Delta$
  propagator is used --- this is discussed below. 
\item{b)}
  The present relativistic and DBT relativistic calculations are in essential
  agreement at $q=550$ MeV/c and up to the $\Delta$ peak for $q=1140$ MeV/c.
\item{c)}
  In the dip following the $\Delta$ peak and above it up to the light cone, our
  transverse response is larger by about a factor two-to-three with respect to
  the DBT results.
\item{d)}
  At variance with the non-relativistic case, the binding energy not only
  shifts the relativistic  $R_T$ toward higher energies, but also
  \textit{increases} it.
\item{e)} Roughly speaking two effects are especially important in
  determining the characteristic behavior seen in Fig.~\ref{fig:RTr}. One is
  the peaking produced by the dynamic $\Delta$ propagation and the second is
  the basic phase space available to the two-nucleon ejection process.
  For instance, the strong rise of $R_T$ close to the light-cone at
  $q=1140$~MeV/c comes from the growth of the available phase space. We have
  checked this by setting all of the current matrix elements entering in 
  Eq.~(\ref{eq:RTDek}) to unity and finding qualitatively the same behavior as
  in the full calculation. 
\end{description}

\begin{figure}
\includegraphics[clip,height=5cm]{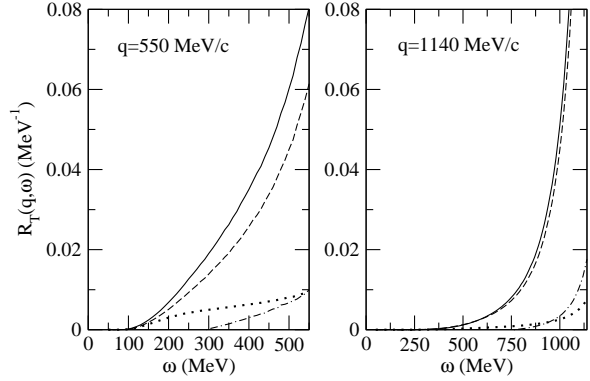}
\caption{\label{fig:RTnrsep} Separate contributions to the transverse response
  function $R_T(q,\omega)$ in the non-relativistic limit at $q=550$~MeV/c and
  $q=1140$~MeV/c: pionic (dotted), pionic-$\Delta$ interference (dash-dotted),
  $\Delta$ (dashed) and total (solid); $k_F=1.3$~fm$^{-1}$. The exchange
contribution is disregarded here.}
\end{figure}

\begin{figure}
\includegraphics[clip,height=5cm]{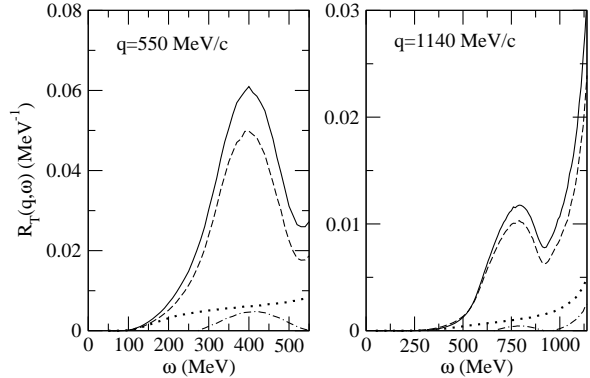}
\caption{\label{fig:RTrsep} As in Fig.~\protect{\ref{fig:RTnrsep}}, but in the
  relativistic case.}
\end{figure}

To gain deeper insight into the transverse response, we display
its individual contributions in Fig.~\ref{fig:RTnrsep} (for the
non-relativistic case) and  in Fig.~\ref{fig:RTrsep} (for the
relativistic one). In both instances the contribution of the
$\Delta$ to $R_T$ is seen to be overwhelming (although this
$\Delta$-dominance is mildly reduced by relativity), the more so
at the larger momentum transfer. This feature strongly suggests
the need for an appropriate treatment of the $\Delta$ degrees of
freedom in obtaining EM responses over the whole kinematical
regime under study. In particular the dynamical propagation of the
latter appears to be essential, as discussed below. 
This in turn implies that a realistic accounting of the width of the $\Delta$
should be undertaken: here, again for ease of comparison with DBT, we have
treated this problem as they did.
Specifically, we have included the width  through the replacement
$M_\Delta\to M_\Delta-i\Gamma(s)/2$ in the energy denominator of
Eq.~(\ref{eq:Deltaprop}), \textit{but not} in the spinor factor.
As far as $\Gamma(s)$ is concerned, $s$ being the square of the
energy of the decaying $\Delta$ in its rest frame, we have adopted
the same expression as in DBT.

\begin{figure}
\includegraphics[clip,height=5cm]{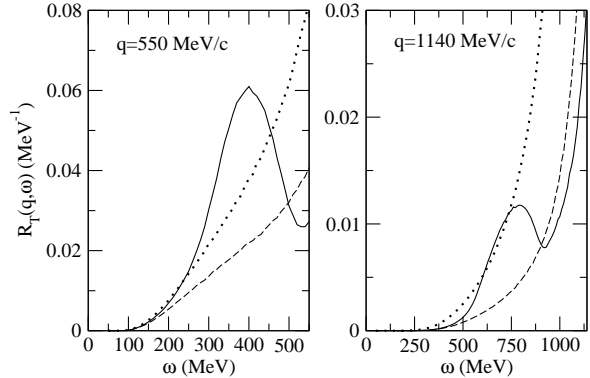}
\caption{\label{fig:RTrprop} The relativistic transverse response function
  $R_T(q,\omega)$ at $q=550$~MeV/c and $q=1140$~MeV/c computed with various
  versions of the $\Delta$ propagator to illustrate the sensitivity to the
  latter: exact propagator (solid), static propagator (dashed) and constant
  propagator (dotted) (see text for the related definitions). In all instances
  $\bar{\epsilon}_2=70$~MeV and $k_F=1.3$~fm$^{-1}$.}
\end{figure}

To assess the crucial role played in $R_T$
by the $\Delta$ propagator more fully, in Fig.~\ref{fig:RTrprop} we
display the relativistic  transverse response
(hence computed with the dynamical $\Delta$ propagator) together with those
obtained by neglecting the frequency dependence of
$S^{\beta\gamma}(p, M_{\Delta})$ (the width of the  $\Delta$ being neglected
as well) in the two possible choices of a static ($p_0=0$) and of a constant
 $\Delta$ propagator. The extreme sensitivity of $R_T$ to
$S^{\beta\gamma}(p, M_{\Delta})$ is clearly shown by the results
in the figure.

One can understand even better the importance of treating correctly the
$\Delta$ propagation by comparing the full relativistic results with a 
calculation employing the non-relativistic $\gamma N\Delta$ and $\pi N\Delta$
vertices, but retaining the relativistic, dynamical energy dependence in the 
denominator of the $\Delta$ propagator, $p^2-M_\Delta^2+iM_\Delta\Gamma$. 
In other words, one can invoke a hybrid model in which everything is kept
non-relativistic, but where the dynamic dependence in the $\Delta$ propagator
is retained. The results are mixed. If one wishes to study only the region
extending through the peak seen in Fig.~\ref{fig:RTrprop} out to the
light-cone {\em and} wishes only to work at relatively small momentum
transfers such as those in the figure, then this hybrid approach is
reasonably good, incurring errors typically of $10\div15\%$.

However, this is not a general statement. First, if one addresses higher
$q$-values, much larger relativistic effects are apparent and the hybrid
approach is not very successful. For instance, at $q=2$($3$)~GeV/c the
typical error made at the peak is $\approx60(100)\%$. Secondly, if one
focuses on the scaling region --- as we shall in the follow-up study that is
presently in progress --- then even at relatively low values of $q$ the
effects of relativity (i.~e., effects other than the nature of the $\Delta$
propagator) are large. For instance, even at $q=1140$~MeV/c one finds effects
exceeding 100\% in the scaling region.
This is not apparent in the figures presented here, since the cross sections
are so small in that region. However, this does not mean that the scaling 
region is irrelevant  --- quite the opposite, the total cross section is also
very small (but measurable) and the 2p-2h MEC effects are far from
negligible. 
The details will be presented in the near future in a separate paper. 

Next let us comment on several general features that characterize
the response in the 2p-2h sector in contradistinction with the
1p-1h case. In particular, the clean theoretical separation
between the 1p-1h and the 2p-2h contributions holds only in the
pure RFG framework. Here, moreover, the 1p-1h response is strictly
confined to the well-known kinematical domain. In finite nuclei,
or even in a correlated infinite Fermi system, neither of the
above statements is valid any longer. We stick however to the pure
RFG because this is the only model that allows us to maintain the
fundamental principles of Lorentz covariance, gauge invariance and
translational invariance.

The 2p-2h response function resides on the entire spacelike domain
of the $(\omega,q)$-plane, in contrast to the 1p-1h case, which,
as mentioned above, is constrained to a limited region --- for
recent fully relativistic studies of the 1p-1h response see
\cite{Ama02,Ama02a,Ama03}. Since momentum conservation involves
only the total momentum there is no limitation on the momentum of
a particle taking part in the 2p-2h excitations and hence the
energy of the latter will vary over a broad range extending from
$\omega=0$ to the light-cone.

In summary, naively the 2p-2h response function $R_T({q},\omega)$ might not
be expected to display any particular structure, but rather to appear as
a broad background. As a matter of fact, as we have seen,
$R_T({q},\omega)$ indeed displays structure, but related not to the nuclear
spectrum involved but to the role played by the $\Delta$ resonance.
Furthermore, we have found that it grows with $\omega$, reflecting the
associated growth of the phase space. Indeed the density per unit
energy of the 2p-2h excitations goes as  $\omega^3$. 

\subsection*{Exchange Contributions}

Considering the purely pionic MEC currents, it is well
known~\cite{Alb90} that the 1p-1h excited states of the RFG are reached
only through the \textit{exchange} term of the related MEC diagrams.
In contrast the 2p-2h states are reached only through the \textit{direct} term.
The origin of these kinds of ``selection rules'' stems, in the first case, from the
spin-isospin saturation of the RFG, in the second from the impossibility
of fulfilling charge conservation in the exchange contribution for the
diagrams displayed in Fig.~\ref{fig:MECpion}.

The $\Delta$-current, however, eludes the above constraints. It provides a
direct contribution in the 1p-1h sector of the RFG spectrum, as well as an
exchange contribution in the 2p-2h sector, both alone and through the
interference with the pionic currents.

\begin{figure}
\includegraphics[clip,height=5cm]{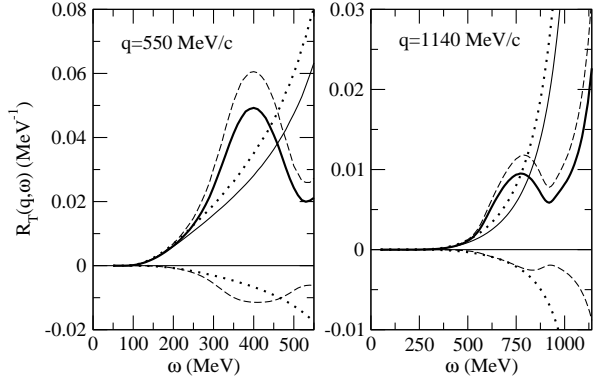}
\caption{\label{fig:RTex} The transverse response function $R_T(q,\omega)$ at
  $q=550$~MeV/c and $q=1140$~MeV/c including the exchange contributions:
  non-relativistic direct (positive dotted), non-relativistic exchange
  (negative dotted), non-relativistic total (light solid), relativistic direct
  (positive dashed), relativistic exchange (negative dashed) and relativistic
  total (heavy solid). In all instances $\bar{\epsilon}_2=70$~MeV and
  $k_F=1.3$~fm$^{-1}$.}
\end{figure}

The exchange contributions are included together with the (larger) direct ones
in Fig.~\ref{fig:RTex}. Two features are immediately apparent
from the figure. First, the impact of relativity on the exchange terms is as
substantial as it is for the direct ones. Moreover, as for the direct case, the
exchange response is first enhanced in the region of the $\Delta$ and then
dampened by the relativistic effects as one approaches the light cone.

Second, the relative magnitude of the exchange terms is seen to
increase both with the energy transfer and with the momentum
transfer. The fraction of the total relativistic response due to
exchange goes from roughly 4\% at low $\omega$ to 29\% at the
light cone for $q=550$~MeV/c and from 7\% to 39\% for
$q=1140$~MeV/c. Thus, the exchange contribution, while not the
dominant one in the kinematical situations we have explored,
appears to us not only far from negligible, but also potentially
interesting in connection with scaling (see below).

Finally, as for the direct case, note that the present calculation
of the non-relativistic exchange contribution agrees with the
calculation at lower momenta reported in ~\cite{Van80}, whereas
does not agree with that of DBT. The relativistic response, on the
other hand, roughly agrees with DBT.

\section{Conclusions}

In this paper we have carried out a fully relativistic calculation
of the MEC contribution of 2p-2h excitations to the transverse
inclusive response within the context of the RFG. These
excitations are reached through the action of the MEC including
the part associated with the $\Delta$ degrees of freedom. We have
compared our results with those of  DBT, the only existing
relativistic study before the present one and we have pointed out
similarities and differences between the two studies. To gauge the
difficulty of this project it might help to stress again, as
anticipated in the Introduction, that the number of terms we have
computed to get $R_T$ amounts to over $100,000$.

Several new aspects contained in the present study bear
highlighting. In particular, while the fully relativistic results
obtained here roughly agree with those reported by DBT, some
substantial differences occur, especially as the lightcone is
approached. Furthermore, our non-relativistic results are
completely in accord with our own previous work; however, they are
not in agreement with DBT in many circumstances. Indeed, the
strongest disagreements occur at high inelasticity and suggest to
us that one should exercise extreme caution when invoking
traditional non-relativistic approximations in such a regime.

Going beyond the extreme non-relativistic approximation \cite{Sch01}, one 
can invoke various classes of hybrid modeling where some, but not all, of the
features of the fully relativistic model are retained
\cite{Ryc01,Mac93,Bof91}. We have pursued such an approach in
which the dynamic $\Delta$ propagator is kept, but otherwise everything
is taken to be non-relativistic. The results are interesting: at modest $q$
(say 1~GeV/c or less) and at high $\omega$ the hybrid approach is quite good,
incurring only $10\div15\%$ error. However, at higher $q$-values, even in the
high $\omega$ region, and at low $\omega$, even for modest $q$, the hybrid
approach is not very successful and relativistic effects are important.

The high $\omega$ region, including the peak produced by the dynamic $\Delta$
propagator and the rise seen as the light-cone is approached when $q$ is
large, might pose some problems for high-energy photoreactions.
The latter are related in the familiar way (see, for example, \cite{DeF66})
to the inclusive electron scattering transverse response by
\begin{equation}
  \sigma^{\text{tot}}_\gamma = 2\pi^2 \frac{\alpha}{\omega} R_T(q,q), 
\end{equation}
  $\alpha$ being the fine-structure constant.

In context, it is important to note that, while one cannot go from these
results directly to studies of (e,e'NN) or ($\gamma$,NN) reactions (see,
e.~g., Refs.~\cite{Car94,Ryc98,Giu01}), since the
RFG is not well-suited to modeling such semi-inclusive processes, the
understanding gained in the present work concerning relativistic effects,
$\Delta$ propagator effects, etc., likely can have an impact, for instance 
via improved descriptions of current operators for use in other approaches. 
The $\Delta$ propagator provides just one example: it produces a prominent
peak in the 2p-2h transverse response and apparently is an important
contribution in the total cross section. It is also very dependent on the
specifics assumed, such as what model is taken for the propagator, how
medium effects are or are not incorporated, how off-shellness is handled, etc. 
In particular, in view of continuing discussions about the off-shell behavior
of the $\Delta$ inside the nuclear medium \cite{Pas98} and of our findings in
this work, the 2p-2h transverse response may provide an ideal place in which
to explore such issues. 

Clearly, this is not the intent of the present work. Here we wish only to
make contact with the work of DBT, using exactly their ingredients and
only in future work will we return to address some of these interesting
issues. 

In particular, in the near future our intent is to address the difficult issue
of the gauge invariance as was explored in the 1p-1h sector in
~\cite{Ama02,Ama02a,Ama03}. This requires the introduction of
2p-2h correlation currents in addition to the MEC. Only when this
task is accomplished will a consistent (namely, gauge invariant),
Lorentz covariant and translational invariant description of the
inclusive responses (both longitudinal and transverse) within the
RFG framework be achieved.

Furthermore, we are now in a position to study the degree to which
these responses fulfill or violate
$y$-scaling of first type (namely in the momentum transfer $q$) and
of second type (namely in $k_F$) --- see \cite{Don99,Don99a,Mai02}.
An exploration of these issues is currently being undertaken for the 2p-2h MEC
and will be presented in a forthcoming paper,
which will also focus on the important theme of the interplay between the
scaling violations arising from the MEC action in both 1p-1h and 2p-2h
sectors. 

\section{Acknowledgments}
This work has been supported in part by the INFN-MIT Bruno Rossi Exchange
Program and by MURST and in part (TWD) by funds provided by
the U.S. Department of Energy under cooperative research agreement
No. DE-DC02-94ER40818.

\appendix*

\section{}
For completeness, in this appendix we report
some of the detailed formulae used for the relativistic calculation.

We start by giving the explicit expressions for Eqs.~(\ref{eq:jcurr}a-b)
which enter in the relativistic $\Delta$-current:
\begin{subequations}
\begin{eqnarray}
\label{eq:jmua}
  j_{(a)}^{\mu}(p,k,q) & = &
       \left\{
       \frac{1}{3M_\Delta^2}\Bigl(
       4\, (k \cdot p) \, p^2 \,\gamma^\mu \rlap/q
      -8\,(k \cdot p)\, (q\cdot p)\, \gamma^\mu \rlap/{p}
      +8\,(k \cdot p) \, p_{\mu}\,\rlap/q \rlap/{p}
       \Bigr.\right. \nonumber \\
      && \Bigl. \;\;\;\;\;\;\;\;\;\;
      - p^2 \rlap/{k}\gamma^\mu \rlap/q \rlap/{p}
      -4\, q^\mu \, (k \cdot p) p^2 \, {\bm 1}
      +\, q^\mu \, p^2 \, \rlap/{k}\rlap/{p}
              \Bigr) \nonumber \\
       && + \frac{2}{3M_\Delta}\Bigl(
       4\, (k \cdot p) \, \gamma^\mu \rlap/q\rlap/{p}
      - p^2\,\rlap/{k}\gamma^\mu \rlap/q
      +2\, (q\cdot p) \,\rlap/{k}\gamma^\mu \rlap/{p}
       \Bigr. \nonumber \\
      && \Bigl. \;\;\;\;\;\;\;\;\;\;
      -2 \, p_{\mu}\,\rlap/{k}\rlap/q\rlap/{p}
      -4\, (k \cdot p) \, q^\mu \,\rlap/{p}
      + q^\mu \,p^2\,\rlap/{k}
              \Bigr) \nonumber \\
      && +     \Bigl(
      \rlap/{k} \gamma^\mu \rlap/q \rlap/{p}
      -2\, \rlap/{p} \gamma^\mu \rlap/q \rlap/{k}
      -3\, q^\mu \, \rlap/{k} \rlap/{p}
      +4\, q^\mu \, (k \cdot p)  \, {\bm 1} \Bigr) \nonumber \\
      && \left. +  2M_\Delta \Bigl(
      -\gamma^\mu \rlap/q \rlap/{k} +
       q^\mu \, \rlap/{k}\Bigr)
      \vphantom{\frac{1}{3M_\Delta^2}}
       \right\} \frac{1}{ (p^2-M_\Delta^2)}, \\
  j_{(b)}^{\mu}(p,k,q) &=&
    \gamma_0 \Bigl(j_{(a)}^{\mu}(p,k,q) \Bigr)^\dagger \gamma_0
\label{eq:jmub}
\end{eqnarray}
\end{subequations}
Useful relations follow from momentum conservation:
$q^\mu+k_1^\mu+k_2^\mu = 0$, $\bm{k}_{1T}+\bm{k}_{2T}=0$ and
$q=q_L= -k_{1L}-k_{2L}$. We also reiterate the definitions:
$p_a\equiv p_1-q$, $p_b\equiv p_1'+q$, $p_c\equiv p_2-q$, $p_d\equiv p_2'+q$,
$k_i\equiv p_i'-p_i$.

Next, in the following formulae, we quote explicitly the integrand entering
in the direct contributions to the transverse response, limiting ourselves to
the full pionic contribution and to the pionic-$\Delta$ interference,
since the pure $\Delta$ contribution corresponds to a far too lengthy
expression to be reported here. We use the notation
${\cal E} \equiv E_{\bm{p}_1}E_{\bm{p}_1'}E_{\bm{p}_2}E_{\bm{p}_2'}$.
The hadronic form factors are not explicitly displayed.

Pionic contribution:
\begin{eqnarray}
{\cal R}_T^{D\pi}(k_1,k_2;q) &=& \frac{V^4}{16{\cal E}} \sum_{\sigma\tau}
    \sum_{ij} \left(\delta_{ij}-\frac{q_i q_j}{\bm{q}^2}\right)
    {J}_i^{\pi\dagger}(k_1,k_2)
    {J}_j^{\pi}(k_1,k_2) \nonumber \\
     &=& \frac{V^4}{16{\cal E}}
   \sum_{\sigma\tau}\sum_{m=x,y}
    {J}_m^{\pi\dagger}(k_1,k_2)
    {J}_m^{\pi}(k_1,k_2) \nonumber \\
   &=& 32 \, \frac{M^4}{\mu_\pi^4{\cal E}} \, \left\{
    {f_{\pi NN}^4 f_{\gamma \pi\pi}^2} \,
    \frac{4\bm{k}_{1T}^2 \, k_1^2 \, k_2^2\; }
    {(k_1^2-\mu_\pi^2)^2(k_2^2-\mu_\pi^2)^2}  \right. \nonumber \\
  && \hspace*{-10mm} \left.
    +{f_{\pi NN}^2 f_{\gamma\pi NN}^2} \, \left[
    \frac{2\,\bm{k}_{1T}^2}{(k_1^2-\mu_\pi^2)(k_2^2-\mu_\pi^2)} -
    \left( \frac{k_1^2(M^2+p_{20}p_{20}'-p_{2L}p_{2L}')}
    {M^2(k_1^2-\mu_\pi^2)^2} + (1\leftrightarrow2)\right) \right] \right.
    \nonumber \\
  && \left. - {f_{\pi NN}^3 f_{\gamma\pi NN}f_{\gamma\pi\pi}}
    \left[ \frac{4\bm{k}_{1T}^2 \, k_1^2}{(k_1^2-\mu_\pi^2)^2(k_2^2-\mu_\pi^2)}
    + (1\leftrightarrow2) \right] \right\};
\label{eq:Rpdir}
\end{eqnarray}

$\Delta$-pion interference contribution:
\begin{eqnarray}
  && {\cal R}_T^{D\{\Delta\pi\}}(k_1,k_2;q) =
    \frac{V^4}{16{\cal E}}\,\sum_{\sigma\tau}\sum_{m=x,y}
    \Bigl[{J}_m^{\Delta\dagger}(k_1,k_2) {J}_m^{\pi}(k_1,k_2) +
    {J}_m^{\pi\dagger}(k_1,k_2) {J}_m^{\Delta}(k_1,k_2) \Bigr] \nonumber \\
  &=& \frac{16M^2 f_{\pi N\Delta}f_{\gamma N\Delta}}{3\mu_\pi^4{\cal E}}
    \left\{ f_{\pi NN}^3 f_{\gamma\pi\pi} \left[
    \frac{-k_2^2\sum\limits_m\Bigl[k_1^m \Bigl({\mathcal{F}}_{a}^m(p_1,p_1';q)-
    {\mathcal{F}}_{b}^m(p_1,p_1';q)\Bigr) \Bigr]}
    {(k_1^2-\mu_\pi^2)(k_2^2-\mu_\pi^2)^2}
    + (1\leftrightarrow 2) \right] \right. \nonumber\\
  &+& \left. f_{\pi NN}^2f_{\gamma\pi NN} \left[
    \vphantom{\frac{2M\sum\limits_m k_{2}^m \Bigl(\Bigr)}{(k_1^2-\mu_\pi^2)}}
    \frac{k_2^2 \Bigl( {\mathcal{S}}_{a}(p_1,p_1';q)
    + {\mathcal{S}}_{b}(p_1,p_1';q) \Bigr)} {4M(k_2^2-\mu_\pi^2)^2}
    - \frac{ \sum\limits_m k_{2}^m \Bigl( {\mathcal{F}}_{a}^m(p_1,p_1';q) +
    {\mathcal{F}}_{b}^m(p_1,p_1';q)\Bigr)}{2(k_1^2-\mu_\pi^2)(k_2^2-\mu_\pi^2)}
    \right.\right. \nonumber \\
  && \left.\left. \hspace*{12cm} + (1\leftrightarrow 2)
    \vphantom{\frac{\sum\limits_m k_{2}^m
    \Bigl({\mathcal{F}}_{a}^m(p_1,p_1';q)\Bigr)}{2(k_1^2-\mu_\pi^2)}}
    \right] \right\};
\label{eq:RDpdir}
\end{eqnarray}

pure $\Delta$ contribution:
\begin{eqnarray}
\label{eq:deltaccia}
  {\cal R}_T^{D\Delta} (k_1,k_2;q) &=&
    \frac{V^4}{16{\cal E}}\sum_{\sigma\tau}\sum_{m=x,y}
    \left( J_m^{\Delta\dagger}(k_1,k_2)J_m^{\Delta}(k_1,k_2)\right)\nonumber \\
  &=& \frac{f_{\pi NN}^2f_{\pi N\Delta}^2f_{\gamma N\Delta}^2}
    {3\mu_\pi^4{\cal E}} \sum_{m=x,y} \left\{\left[\left(\vphantom{\frac{1}{1}}     \frac{{\mathcal{F}}_{a}^m(p_1,p_1';q) {\mathcal{F}}_{b}^m(p_2,p_2';q)}
    {3(k_1^2-\mu_\pi^2)(k_2^2-\mu_\pi^2)} \right. \right. \right.\nonumber \\
  && + \frac{-k_2^2}{(k_2^2-\mu_\pi^2)^2} \Bigl[ \bm{Tr}\left\{
    (\rlap/{p_1}-M) j_{(b)}^m(p_a,k_2,q) (\rlap/{p_1}'+M) j_{(a)}^m(p_a,k_2,q)
    \right\} \nonumber \\
  &&  \hspace*{25mm} + \bm{Tr}\left\{
    (\rlap/{p_1}+M) j_{(a)}^m(p_b,k_2,q) (\rlap/{p_1}'-M)
    j_{(b)}^m(p_b,k_2,q) \right\} \nonumber \\
  && \Bigl. \left. \left. - \frac{2}{3} \bm{Tr} \left\{
    (\rlap/{p_1}+M) j_{(a)}^m(p_b,k_2,q) (\rlap/{p_1}'-M)
    j_{(a)}^m(-p_a,k_2,q) \right\} \Bigr]\right)
  + (1\leftrightarrow 2)\right] \nonumber \\
  && \left.-\frac{{\mathcal{F}}_{a}^m(p_1,p_1';q){\mathcal{F}}_{a}^m(p_2,p_2';q)    +{\mathcal{F}}_{b}^m(p_1,p_1';q){\mathcal{F}}_{b}^m(p_2,p_2';q)}
    {3(k_1^2-\mu_\pi^2)(k_2^2-\mu_\pi^2)}\right\}.
\end{eqnarray}
In the previous equations the following definitions have been introduced:
\begin{subequations}
\begin{eqnarray}
\label{eq:RDapfdir}
  {\mathcal{F}}_{a}^m(p,p';q) &\equiv&
    \bm{Tr}\left\{ (\rlap/{p}-M)(\rlap/{p}'+M)
    j_{(a)}^m(p-q , -q-p'+p, q) \right\} \nonumber \\
  &=& \bm{Tr}\left\{ (\rlap/{p}-M)(\rlap/{p}'+M)
    j_{(a)}^m(p_\Delta,k_\pi,q)\right\} \nonumber \\
  &=& \frac{1}{p_\Delta^2-M_\Delta^2}
    \left\{ \vphantom{\begin{array}{c}a \\ a \end{array}}
    \frac{4 q^2}{3M_\Delta^2}
    \Bigl[ p'^m (p\cdot q) - p^m(p'\cdot q)+ 2p^m (p\cdot p') \Bigr]
    \Bigl[  p_\Delta^2-4(k_\pi\cdot p_\Delta)  \Bigr] \right. \nonumber \\
  && + \frac{4M^2}{3M_\Delta^2} \left\{ 4(k_\pi\cdot p_\Delta)
    \Bigl[ p^m(p'\cdot q)-p'^m (p\cdot q) +2p'^m q^2 \Bigr]
    -q^2  p_\Delta^2 (p^m+p'^m)\right\} \nonumber \\
  && + \frac{8M}{3M_\Delta}
    \left\{  k_\pi^m \Bigl[ 4(k_\pi\cdot p_\Delta)(q\cdot p_\Delta)
    -p_\Delta^2 q^2+2(q\cdot p_\Delta)^2 \Bigr] \right.\nonumber \\
  && \left. \hspace*{15mm}
    -2p^m\Bigl[q^2(k_\pi\cdot p_\Delta)+(k_\pi\cdot q)(q\cdotp_\Delta )
    +2 (k_\pi\cdot q) (k_\pi\cdot p_\Delta) \Bigr]\right\} \nonumber \\
  && +4 \left\{ q^2\Bigl[p'^m (q\cdot p)-p^m(q\cdot p')\Bigr]
    +2 (p\cdot p')\Bigl[p^m q^2+2 p^m (q\cdot p')-2p'^m (q\cdot p_\Delta)\Bigr]
    \right\} \nonumber \\
  && \left. + 4M^2 \left\{ 4 p'^m \Bigl[(q\cdot p_\Delta) -p^m(p'\cdot q)\Bigr]
    -q^2(p'^m+p^m) \right\} + 8 M M_\Delta  k_\pi^m  q^2
    \vphantom{\begin{array}{c}a \\ a \end{array}}
    \right\}, \\
\label{eq:RDbpfdir}
  {\mathcal{F}}_{b}^m(p,p';q) &\equiv &
    \bm{Tr}\left\{ (\rlap/{p}+M)(\rlap/{p}'-M)
    j_{(b)}^m(p'+q , -q-p'+p, q) \right\} \nonumber \\
  &=& \bm{Tr}\left\{ (\rlap/{p}'-M)(\rlap/{p}+M)
    j_{(a)}^m(p'+q , -q-p'+p , q) \right\} =
    {\mathcal{F}}_{a}^m(p',p;-q).
\end{eqnarray}
\end{subequations}
\begin{subequations}
\begin{eqnarray}
  {\mathcal{S}}_{a}(p,p';q) &\equiv& \sum_{m=x,y}
    \bm{Tr}\left\{ (\rlap/{p}-M)\gamma^m(\rlap/{p}'+M)
    j_{(a)}^m(p-q,-q-p'+p,q)\right\} \nonumber \\
  &=& \sum_{m=x,y}  \bm{Tr}\left\{ (\rlap/{p}-M)\gamma^m(\rlap/{p}'+M)
    j_{(a)}^m(p_\Delta,k_\pi,q)\right\} \nonumber \\
  &=& \left\{ \vphantom{\begin{array}{c}a \\ a \end{array}}
    \frac{8M}{3M_\Delta^2} \left\{ 4 (k_\pi\cdot p_\Delta) \Bigl[
    2 (q\cdot p_\Delta) \Bigl((p'\cdotp_\Delta) -(q\cdot p1)\Bigr)
    -p_\Delta^2 (q\cdot p1) \right.\right.\Bigr.\nonumber \\
  && \left. \left. \Bigl. -\bm{p}_T^2\Bigl(q^2+(q\cdot pp1)\Bigr)
    +\bm{p}_T\cdot\bm{p}_T(q\cdot p_\Delta)\Bigr]
    +p_\Delta^2 \Bigl[q^2(k_\pi\cdot p) -(q\cdot k_\pi)(p'\cdot p_\Delta)
    \right.\right.\Bigr. \nonumber \\
  && \left.\Bigl. -3(q\cdot p')(k_\pi\cdot  p_\Delta)
    -(q\cdot p_\Delta)(k_\pi\cdot p') +
    \bm{k}_{\pi T}\cdot\bm{p}_T \Bigl(q^2+(q\cdot p')\Bigr)
    \right. \nonumber \\
  && \left. - \bm{p}_T\cdot\bm{p}_T'(q\cdot k_\pi)\Bigr] \right\}
    + \frac{8M^3}{3M_\Delta^2} \Bigl\{8(q\cdot p_\Delta)(k_\pi\cdot p_\Delta)
    -p_\Delta^2 (q\cdot k_\pi)\Bigr\} \nonumber \\
  && + \frac{16q^2}{3M_\Delta}
    \Bigl\{(q\cdot p')(k_\pi\cdot p) - (q\cdot p)(k_\pi\cdot p')
    - (q\cdot k_\pi)(p\cdot p') -4 (p_\Delta\cdot k_\pi)(p\cdot p')
    \Bigr. \nonumber \\
  && \Bigl. + \bm{p}_T\cdot\bm{p}_T'
    \Bigl[p_\Delta^2-(p_\Delta\cdot k_\pi)\Bigr] \Bigr\}
    +\frac{16M^2}{3M_\Delta}
    \Bigl\{ (q\cdot k_\pi)\Bigl[(p\cdot p')+p_\Delta^2\Bigr]
    -(q\cdot p)(k_\pi\cdot p') \Bigr. \nonumber \\
  && \left. +2q^2(k_\pi\cdot p') +3(q\cdot p')(k_\pi\cdot p)
    -4(q\cdot p')(k_\pi\cdot q) +2(q\cdot p_\Delta)(k_\pi\cdot p_\Delta)
   \right.\nonumber \\
  && \left. -\bm{p}_T^2\Bigl(q^2+(q\cdot p')\Bigr)+
    \bm{p}_T\cdot\bm{p}_T'(q\cdot p_\Delta)\Bigr\}
    +8M \Bigl\{ q^2 (k_\pi\cdot p) -(q\cdot k_\pi)(p'\cdot p_\Delta)
    \right.\Bigr. \nonumber \\
  && \Bigl.\left. +(q\cdot p')(k_\pi\cdot p_\Delta) -
    (q\cdot p_\Delta) (k_\pi\cdot p')
    +\bm{k}_{\pi T}\cdot\bm{p}_T(q\cdot p')
    -2\bm{k}_{\pi T}\cdot\bm{p}_T'(q\cdot p_\Delta)
    \right.\Bigr. \nonumber \\
  && \Bigl.\left. +\bm{p}_T^2\Bigl[q^2+2(q\cdot p')\Bigr]
    -\bm{p}_T\cdot\bm{p}_T'\Bigl[(q\cdot p)+(q\cdot p')\Bigr]
    \Bigr\}-8M^3\Bigl\{(q\cdot k_\pi)\Bigr\} \right.\nonumber \\
  && \left. +16M_\Delta \Bigl\{(p\cdot p')\Bigl[q^2+2(q\cdot p')\Bigr]
    +\bm{p}_T\cdot\bm{p}_T'\Bigl[q^2+(q\cdot p')\Bigr]
    -\bm{p}_T'^2(q\cdot p)\Bigr\} \right. \nonumber \\
  && \left. +16M_\Delta M^2\Bigl\{q^2-2(q\cdot p)\Bigr\}
    \vphantom{\begin{array}{c}a \\ a \end{array}} \right\}
    \frac{1}{p_\Delta^2-M_\Delta^2} \nonumber \\
\label{eq:RDasdir}
\end{eqnarray}
and
\begin{eqnarray}
\label{eq:RDbsdir}
  {\mathcal{S}}_{b}(p,p';q) &\equiv& \sum_{m=x,y}
    \bm{Tr}\left\{ (\rlap/{p}+M)\gamma^m(\rlap/{p}'-M)
    j_{(b)}^m(p'+q,-q-p'+p,q)\right\} = {\mathcal{S}}_{a}(p',p;-q).
\nonumber \\
\end{eqnarray}
\end{subequations}
In Eqs.~(\ref{eq:RDapfdir}) and (\ref{eq:RDasdir}) we used $p_\Delta=p-q$ and
$k_\pi=-q-p'+p$  to simplify the expressions.
Eqs.~(\ref{eq:RDbpfdir}) and (\ref{eq:RDbsdir}) are consequences
of the identities: $j_{(a,b)}^m (p,k,q)= j_{(a,b)}^m (p,-k,-q)$ and
Eq.~(\ref{eq:jmub}).

The traces indicated in Eq.~(\ref{eq:deltaccia}) involve thousands of
terms and are not included here.

From Eqs.~(\ref{eq:Rpdir})--(\ref{eq:deltaccia})
the non-relativistic limit of Eq.~(\ref{eq:MECint}) can be obtained by assuming
that all the nucleonic momenta are much smaller than the nucleon rest mass.

\end{document}